\documentclass[onecolumn,secnumarabic,amssymb, amsmath, nofootinbib,tightenlines,
nobibnotes, aps, prl,epsfig, pdffig]{revtex4}
\usepackage{graphicx}
\usepackage{color}
\usepackage{dcolumn}
\usepackage{bm}
\begin{document}
\preprint{APS/123-QED}
\title{Importance of the heavy-quarks longitudinal structure function measurements  at future circular collider energies }

\author{G.R.Boroun}%
 \email{grboroun@gmail.com; boroun@razi.ac.ir }

\affiliation{ Physics Department, Razi University, Kermanshah
67149, Iran}
\date{\today}
\begin{abstract}
In this article, we consider the ratio of
 structure functions for heavy quark pair production at low values of $x$. The importance of this ratio for
charm and beauty pair production is examined according to the
Hadron Electron Ring Accelerator (HERA) data. The behavior of
these ratios  considers due to the hard-pomeron behavior of the
gluon distribution function. The results are in good agreement
with respect to the HERA data. Expanding this data to the range of
new energies underscore the importance of these measurements for
heavy quarks. The ratio of charm and beauty structure functions at
the proposed Large Hadron electron Collider (LHeC) is considered
as a function of invariant center-of-mass energy. For top pair
production this ratio is extracted with known kinematics of the
LHeC and Future Circular Collider electron-hadron (FCC-eh)
colliders. Comparison of the results obtained for the ratio of top
structure functions  in LHeC and FCC-eh are proportional to the
specified
inelasticity $y$ range.\\
\end{abstract}
 \pacs{***}
\maketitle
\subsection{1. Introduction}
The latest data collected in HERA for heavy quarks show that the
cross sections for charm and beauty production can be considered
in a wide range of $x$ and $Q^{2}$ values from the H1 and ZEUS
detectors [1-4]. Also a combination method at HERA used for the
cross section data  with respect to the correlations of the
statistical and systematic uncertainties. In neutral current (NC)
deep inelastic electron-proton scattering (DIS) at HERA,
heavy-quarks production is the most important quantum
chromodynamics (QCD) tests. The dominant process in the production
of heavy quarks is Boson-gluon-fusion (BGF). The production of
heavy quarks at HERA depends on the mass of these quarks and thus
the calculations of cross sections depend on a wide range of
perturbative scales $\mu^{2}$. The massive fixed-flavour-number
scheme (FFNS) [5] and the variable-flavour-number scheme (VFNS)
[6] are different approaches for  considering heavy quarks. FFNS
can be used on the threshold of $\mu^{2}{\approx}m^{2}$ and for
$\mu^{2}{\gg}m^{2}$ VFNS is used where the treatment of
resummation of collinear logarithms $\ln(\mu^{2}/m^{2})$ is
achieved. In Ref.[7] a general-mass variable-flavour-number scheme
(GM-VFNS)   for
calculation of the contributions of heavy quarks introduced.\\
In the process
$\gamma^{*}g{\rightarrow}\mathcal{Q}\overline{\mathcal{Q}}$, where
$\mathcal{Q}$ is heavy quark, heavy-quark production is sensitive
to the gluon distribution and gluon momentum transfer in the
proton. In accordance with the heavy quarks mass, the gluon
momentum is arranged such that $x_{g}^{t}>x_{g}^{b}>x_{g}^{c}$.
HERA dataset, for production charm and beauty in DIS, covers the
kinematic range of photon virtuality $2.5 {\leq} Q^{2} {\leq}
2000~\mathrm{GeV}^{2}$ and Bjorken scaling variable $3.10^{-5}
{\leq} x {\leq} 5.10^{-2}$ [1-4]. The ep center-of-mass energies
with data taken with the H1 and ZEUS detectors are corresponding
to $\sqrt{s}=319~\mathrm{GeV}$ [1] and $\sqrt{s}=318~\mathrm{GeV}$
[2] respectively. Future circular electron proton colliders are
the ideal environment to increase center-of-mass energy [8-10]. At
LHeC, the electron-proton center of mass energy  reach to
$\sqrt{s} \simeq 1.3~ \mathrm{TeV}$. The $x,Q^{2}$ values of
simulated heavy quark density data used in LHeC studies reach to
$x{\leq}10^{-5}$ and $Q^{2}{\geq}10^{4}~\mathrm{GeV}^{2}$ [9].
Also the heavy quark densities will be checked at FCC-eh programme
which the center-of-mass energy
reaches $\simeq 3.5~ \mathrm{TeV}$ [10].\\
The heavy-quark structure functions obtained in DIS at HERA, from
the measurement of the inclusive heavy quark cross sections, are
an important test of the QCD. These structure functions are
obtained after applying small corrections for the heavy-quark
longitudinal structure functions [1-4]. The heavy-quark cross
section defined in terms of the heavy-quark structure functions as
\begin{eqnarray}
\frac{d^{2}\sigma^{\mathcal{Q}\overline{\mathcal{Q}}}}{dxdQ^{2}}
&=&\frac{2{\pi}\alpha^{2}(Q^{2})}{xQ^{4}}\{[1+(1-y)^{2}]F_{2}^{\mathcal{Q}\overline{\mathcal{Q}}}(x,Q^{2})
-y^{2}F_{L}^{\mathcal{Q}\overline{\mathcal{Q}}}(x,Q^{2}) \},
\end{eqnarray}
where $y$ is the inelasticity. The reduced cross section defined
as follows
\begin{eqnarray}
\sigma^{\mathcal{Q}\overline{\mathcal{Q}}}_{\mathrm{red}}&=&\frac{d^{2}\sigma^{\mathcal{Q}\overline{\mathcal{Q}}}}{dxdQ^{2}}
.\frac{xQ^{4}}{2{\pi}\alpha^{2}(Q^{2})(1+(1-y)^{2})}\nonumber\\
&&=F_{2}^{\mathcal{Q}\overline{\mathcal{Q}}}(x,Q^{2})-f(y)F_{L}^{\mathcal{Q}\overline{\mathcal{Q}}}(x,Q^{2}),
\end{eqnarray}
where $f(y)=\frac{y^{2}}{1+(1-y)^{2}}$. In HERA kinematic range
the contribution $F_{L}^{\mathcal{Q}\overline{\mathcal{Q}}}$ is
small. Therefore the heavy-quark structure function
$F_{2}^{\mathcal{Q}\overline{\mathcal{Q}}}$ is obtained from the
measured heavy-quark cross sections. The measurements of
$\sigma^{\mathcal{Q}\overline{\mathcal{Q}}}_{\mathrm{red}}$, based
on data from HERA I and HERA II, are shown as a function of $x$
for fixed values of $Q^{2}$ in Refs.[1-4]. The measured values of
heavy-quark structure functions
$F_{2}^{\mathcal{Q}\overline{\mathcal{Q}}}$ were obtained using
\begin{eqnarray}
F_{2}^{\mathcal{Q}\overline{\mathcal{Q}}}(x,Q^{2})=
\frac{d^{2}\sigma^{\mathrm{jet}}_{\mathcal{Q}}/{dxdQ^{2}}}
{d^{2}\sigma^{\mathrm{had,NLO}}_{\mathcal{Q}}/{dxdQ^{2}}}
F_{2}^{\mathcal{Q}\overline{\mathcal{Q}},\mathrm{NLO}}(x,Q^{2})
\end{eqnarray}
The differential cross section for jet production method is
defined in literature and also NLO QCD predictions were obtained
from the FFNS using HVQDIS program [11]. This method is also used
for the heavy-quark longitudinal structure function to the cross
section. Heavy-quark cross sections were determined and extracted
in analogy to $F_{2}^{\mathcal{Q}\overline{\mathcal{Q}}}$. In this
way no assumption on $F_{L}^{\mathcal{Q}\overline{\mathcal{Q}}}$
was required. Indeed
\begin{eqnarray}
\sigma^{\mathcal{Q}\overline{\mathcal{Q}}}_{\mathrm{red}}
=F_{2}^{\mathcal{Q}\overline{\mathcal{Q}}}(x,Q^{2})[1-f(y)F_{L}^{\mathcal{Q}\overline{\mathcal{Q}}}(x,Q^{2})/F_{2}^{\mathcal{Q}\overline{\mathcal{Q}}}(x,Q^{2})]
\end{eqnarray}
Future circular colliders will extend the ratio
$F_{L}^{\mathcal{Q}\overline{\mathcal{Q}}}/F_{2}^{\mathcal{Q}\overline{\mathcal{Q}}}$
into a region of much smaller $x$ and higher $Q^2$. Indeed the
LHeC is the ideal place to resolve this ratio [9]. Overview of the
kinematic plane of the LHeC pseudo-data [12,13] is illustrated in
Table I for the $x, Q^{2}$ values of simulated heavy quark density
data [9,14]. These data provide additional constrains on the
gluon.\\
The paper is organized as follows. In sect.2, we give a summary
about the ratio of heavy-quark structure functions. Then we
introduce a method to calculate the ratio
$F_{L}^{\mathcal{Q}\overline{\mathcal{Q}}}/F_{2}^{\mathcal{Q}\overline{\mathcal{Q}}}$
applying the gluon exponent. After reviewing the essential
features of the charm and beauty structure functions at HERA in
section 3, we will study the heavy-quark structure functions  at
the LHeC and FCC-eh kinematics in this section.  Section 4 contain
the results and
discussions.  Our summary and conclusion are given in section 5.\\

\subsection{2. A Short Theoretical Input}

The dynamics of flavor-singlet quark and gluon distribution
functions, $q^{s}$ and $g$, are defined by
\begin{eqnarray}
q^{s}(x,n_{f},\mu^{2})&=&\sum_{l=1}^{n_{f}}[f_{l}(x,n_{f},\mu^{2})+\overline{f}_{l}(x,n_{f},\mu^{2})],\nonumber\\
g(x,n_{f},\mu^{2})&=&f_{g}(x,n_{f},\mu^{2}),
\end{eqnarray}
where $n_{f}$ is number of active quark flavor. The heavy-quark
structure functions derived  using the zero-mass VFN scheme
(ZMVFN) as
\begin{eqnarray}
F^{ZMVFN}_{k,\mathcal{Q}}=\sum_{j=0}^{\infty}a_{s}^{j}(n_{f}+1)\sum_{i=q,g,\mathcal{Q}}
C_{k,i}^{(j)}(n_{f}+1){\otimes}f_{i}(n_{f}+1)
\end{eqnarray}
where $C^{,}s $ are the Wilson coefficients at the $j$-th order
and $k=2$ and $L$. Here $a_{s}=\frac{\alpha_{s}}{4\pi}$ is the QCD
running coupling. Eq.(6), at asymptotically large momentum
transfer $Q^{2}{\gg}m_{\mathcal{Q}}^{2}$, is valid. For
$Q^{2}{\simeq}m_{\mathcal{Q}}^{2}$ VFNS is valid which it includes
a combination of the ZMVFN with FFNS. In this case the heavy-
quark structure functions are
\begin{eqnarray}
F^{FFNS}
_{k,\mathcal{Q}}=\sum_{j=0}^{\infty}a_{s}^{j}(n_{f})\sum_{i=q,g}
H_{k,i}^{(j)}(n_{f}){\otimes}f_{i}(n_{f}),
\end{eqnarray}
where $H^{,}s$ are the Wilson coefficients for the DIS heavy-quark
production [15]. In the following, we suppress the dependence on
the active flavor $n_{f}$. In GM-VFNS on should take into account
quark mass as  the rescaling variable $\chi$ is defined into the
Bjorken variable $x$ by the following form [16]
\begin{eqnarray}
\chi=x(1+\frac{4m_{\mathcal{Q}}^{2}}{Q^{2}}),
 \end{eqnarray}
where the rescaling variable, at high $Q^{2}$ values
($m^{2}_{\mathcal{Q}}/Q^{2}\ll 1$), reduces to $x$,
$\chi{\rightarrow}x$. Here $m_{\mathcal{Q}}$ is the heavy quark
mass.\\
Within this scheme  heavy quark densities arise via the
$g{\rightarrow}\mathcal{Q}\overline{\mathcal{Q}}$ evolution. In
the small-$x$ range the heavy quark contributions
 are given by these forms
\begin{eqnarray}
F_{2}^{\mathcal{Q}\overline{\mathcal{Q}}}(x,Q^{2})=C_{2,g}^{\mathcal{Q}\overline{\mathcal{Q}}}(x,\xi){\otimes}G(x,\mu^{2}),\nonumber\\
F_{L}^{\mathcal{Q}\overline{\mathcal{Q}}}(x,Q^{2})=C_{L,g}^{\mathcal{Q}\overline{\mathcal{Q}}}(x,\xi){\otimes}G(x,\mu^{2}).
\end{eqnarray}
where $\xi=\frac{m^{2}_{\mathcal{Q}}}{Q^{2}}$. The coefficient
functions up to NLO analysis, $C_{2,g}$ and $C_{L,g}$,
demonstrated in Refs.[5,15,17]. $\otimes$ note the convolution
between two functions of $x$ as, $[f\otimes
g](x)=\int_{x}^{1}(dy/y)f(y)g(x/y)=\int_{x}^{1}(dz/z)f(x/z)g(z)$.
Here $G(x,\mu^{2})(=xg(x,\mu^{2}))$ is the gluon momentum
distribution and the default renormalisation scale $\mu_{r}$ and
factorization scale $\mu_{f}$ are set to $\mu\equiv
\mu_{r}=\mu_{f}=\sqrt{Q^{2}+4m^2_{\mathcal{Q}}}$. Within the pQCD,
and up to the NLO corrections, the coefficient functions $C_{2,g}$
and $C_{L,g}$ read as
\begin{eqnarray}
C_{k,g}(z,\xi)=C^{0}_{k,g}(z,\xi)+a_{s}(\mu^{2})C^{1}_{k,g}(z,\xi).
\end{eqnarray}
 The behavior of the heavy-quark structure
function is governed entirely by the hard pomeron exchange in
accordance with the Regge theory. The heavy quark cross section
depends strongly on the gluon distribution which the density of
gluons enter at
\begin{eqnarray}
x_{g}^{\mathcal{Q}}=\frac{Q^{2}+M^{2}_{t}}{W^{2}+Q^{2}}=x_{bj}(1+\frac{M^{2}_{t}}{Q^{2}}),
\end{eqnarray}
where $M_{t}$ is the transverse mass of the produced heavy quark
pair $\mathcal{Q}\overline{\mathcal{Q}}$ [18]. The standard
parameterization of the gluon distribution function at low $x$
observed by the following form
\begin{eqnarray}
G(x,Q^{2})_{x{\rightarrow}0}=f_{g}(Q^{2})x^{-\lambda_{g}(Q^{2})}.
\end{eqnarray}
Over a wide range of $Q^{2}$, the gluon density behaves as a fixed
power of $x$ as the quantity $\lambda_{g}$ is the hard-pomeron
intercept minus one \footnote{ Authors in Refs.[19] obtained a
good numerical fit to the output of the DGLAP evolution for the
gluon distribution at low $x$ by the following form $
G(x,Q^{2})=f_{g}(Q^{2})x^{-\epsilon_{0}} $ where $\epsilon_{0}$ is
hard pomeron exchange. Over a wide range of $Q^{2}$ values, the
charm structure function behaves as a fixed power of $x$ as $
F_{2}^{c}(x,Q^{2})=f_{c}(Q^{2})x^{-\epsilon_{0}} $.}. Eq.(12)
granted originates of heavy quarks from gluons in the proton.
Indeed the heavy-quark structure functions can be described by a
fixed power of $x$ behavior [19,20] as
\begin{eqnarray}
F_{2}^{c(b)}=f_{\mathcal{Q}}(Q^{2})x^{-\lambda_{eff}}.
\end{eqnarray}
The gluon exponent at low values of $x$ is described by the
hard-pomeron as the fixed coupling LLx BFKL
  solution gives  the value  $\lambda_{g}{\simeq}-0.5$. Although dependence on $Q^{2}$ is expressed in  effective exponent defined
  in Ref.[21]. In literatures, fit to experimental data suggests that hard pomeron
   dominate the behavior of the heavy quark structure function. It is suggested that this behavior can be shown
   in the form of Eq.(13), which we assume that
   $\lambda_{eff}^{c}{\simeq}\lambda_{eff}^{b}$. Because the heavy quark behavior at low $x$ is determined by the gluon behavior, so
  $\lambda_{eff}^{c}{\simeq}\lambda_{eff}^{b}=\lambda_{g}$.\\
    Within the dipole formulation of the $\gamma^{*} p$ scattering
  [22], the gluon density is modelled as
  $xg(x,\mu^{2})=\frac{3}{4\pi^{2}\alpha_{s}}\frac{\sigma_{0}}{R_{0}^{2}(x)}$,
  where $R_{0}(x)$ is the saturation scale. This function
  decreases when $x{\rightarrow}0$ as
  $R_{0}^{2}(x)=\frac{1}{\mathrm{GeV}^{2}}(\frac{x}{x_{0}})^{\lambda_{g}}$.
  The parameters of the model (i.e., $\sigma_{0}$ and $x_{0}$) are
  defined in Ref.[22]. In color dipole model (CDM) the saturation exponent $\lambda_{_{g}}$ is defined from a fit to low
  $x$ data as $\lambda_{g}{\simeq}0.3$. In this analysis,
   we will try to select the gluon exponent value corresponding to the average of the hard pomeron and color dipole
   model, where $0.3{\leq}\lambda_{g}{\leq}0.5$. Here the lower limit corresponds to the saturation exponent  and the upper limit corresponds to
  the hard pomeron exponent.\\
  In recent years [23,24], the phenomenological various
  successful methods have examined charm and beauty structure
  functions. This importance, along with the t-quark density, can be explored at future circular collider
  energies. One of the important top quark production modes is
  $t\overline{t}$ photoproduction [26,26,27]. The total cross section
  prediction at LHeC is 0.05 pb [28]. These studies lead us to new physics in the
  future.\\

\subsection{3. Method}
Based on the hard-pomeron behavior of the distribution functions,
the ratio of heavy-quark structure functions is formulated based
on the coefficient functions and gluon exponent. After doing the
integration over $z$, the ratio $F_{L}/F_{2}$ for heavy quarks can
be rewritten in a  convolution form as the ratio
$F_{L}^{\mathcal{Q}\overline{\mathcal{Q}}}/F_{2}^{\mathcal{Q}\overline{\mathcal{Q}}}$
is defined by
\begin{eqnarray}
\frac{F_{L}^{\mathcal{Q}\overline{\mathcal{Q}}}}{F_{2}^{\mathcal{Q}\overline{\mathcal{Q}}}}
=\frac{C_{L,g}^{Q\overline{Q}}(x,\xi){\odot}x^{\lambda_{g}}}{C_{2,g}^{Q\overline{Q}}(x,\xi){\odot}x^{\lambda_{g}}},
\end{eqnarray}
where $[f{\odot} g](x)=\int_{x}^{1}(dy/y)f(y)g(y)$. In the
analytical form, power-like behavior of the heavy-quark structure
functions is generically written as follows
\begin{eqnarray}
\frac{\partial}{\partial{\ln}\frac{1}{x}}{\ln}\frac{F_{L}^{\mathcal{Q}\overline{\mathcal{Q}}}(x,Q^{2})}{F_{2}^{\mathcal{Q}\overline{\mathcal{Q}}}(x,Q^{2})}
&=&
\lambda_{L}^{\mathcal{Q}\overline{\mathcal{Q}}}-\lambda_{2}^{\mathcal{Q}\overline{\mathcal{Q}}}\nonumber\\
&&=\frac{\partial}{\partial{\ln}\frac{1}{x}}{\ln}\frac{C_{L,g}^{\mathcal{Q}\overline{\mathcal{Q}}}
(x,\xi){\odot}x^{\lambda_{g}}}{C_{2,g}^{\mathcal{Q}\overline{\mathcal{Q}}}(x,\xi){\odot}x^{\lambda_{g}}}.\nonumber\\
\end{eqnarray}
The exponents $\lambda_{L}$ and $\lambda_{2}$ for heavy-quark
production are defined by the derivatives of the heavy-quark
structure functions by  the following forms
\begin{eqnarray}
 \lambda_{L}^{\mathcal{Q}\overline{\mathcal{Q}}}&=&{\partial \ln F_{L}^{\mathcal{Q}\overline{\mathcal{Q}}}(x,Q^{2})}/{\partial
\ln(1/x)},\nonumber\\
 \lambda_{2}^{\mathcal{Q}\overline{\mathcal{Q}}}&=&{\partial \ln F_{2}^{\mathcal{Q}\overline{\mathcal{Q}}}(x,Q^{2})}/{\partial
\ln(1/x)}.
\end{eqnarray}
The importance of the relationship between
$\sigma^{\mathcal{Q}\overline{\mathcal{Q}}}_{\mathrm{red}}$ and
$F_{2}^{\mathcal{Q}\overline{\mathcal{Q}}}(x,Q^{2})$ in Eq.(4)
depends on the functions of $f(y)$ and the ratio
$F_{L}^{\mathcal{Q}\overline{\mathcal{Q}}}/F_{2}^{\mathcal{Q}\overline{\mathcal{Q}}}$.
In the high inelasticity which $f(y){\rightarrow}1$, the
importance of the longitudinal structure function  in the
production of heavy pair quarks is revealed in the LHeC and
FCC-eh. In comparisons with the latest data collected in HERA [3],
we can see an increase in values of $Q^{2}$ and a decrease in
values of $x$ in new energies.\\
HERA data are expressed in terms of two variables, $x$ and
$Q^{2}$. At low $x$ we define a new variable that
$Q^{2}/x{\simeq}W^{2}$. $W^{2}$ refers to the photon-proton
center-of-mass energy. Indeed the heavy-quark structure functions
are given by the single variable $W^{2}$ as
\begin{eqnarray}
F_{L}^{\mathcal{Q}\overline{\mathcal{Q}}}(x,Q^{2})
=F_{L}^{\mathcal{Q}\overline{\mathcal{Q}}}(W^{2}=\frac{x}{Q^{2}},Q^{2}),\nonumber\\
F_{2}^{\mathcal{Q}\overline{\mathcal{Q}}}(x,Q^{2})
=F_{2}^{\mathcal{Q}\overline{\mathcal{Q}}}(W^{2}=\frac{x}{Q^{2}},Q^{2}).
\end{eqnarray}
According to  power-like behavior, the heavy quark structure
functions can be stated as
\begin{eqnarray}
F_{L}^{\mathcal{Q}\overline{\mathcal{Q}}}(W^{2}){ \sim}
(W^{2})^{\lambda_{L}},\nonumber\\
F_{2}^{\mathcal{Q}\overline{\mathcal{Q}}}(W^{2}){ \sim}
(W^{2})^{\lambda_{2}}.
\end{eqnarray}
The exponents now are defined by the following forms
\begin{eqnarray}
 \lambda_{L}^{\mathcal{Q}\overline{\mathcal{Q}}}&=&\frac{\partial \ln F_{L}^{\mathcal{Q}\overline{\mathcal{Q}}}(W^{2})}{\partial
\ln W^{2}},\nonumber\\
 \lambda_{2}^{\mathcal{Q}\overline{\mathcal{Q}}}&=&\frac{\partial \ln F_{2}^{\mathcal{Q}\overline{\mathcal{Q}}}(W^{2})}{\partial
\ln W^{2}}.
\end{eqnarray}
\begin{eqnarray}
{\Rightarrow}~{\Delta}\lambda_{L2}^{\mathcal{Q}\overline{\mathcal{Q}}}&=&
\lambda_{L}^{\mathcal{Q}\overline{\mathcal{Q}}}-
\lambda_{2}^{\mathcal{Q}\overline{\mathcal{Q}}}\nonumber\\
&&=\frac{\partial }{\partial \ln W^{2}}\ln
\frac{F_{L}^{\mathcal{Q}\overline{\mathcal{Q}}}(W^{2})}{F_{2}^{\mathcal{Q}\overline{\mathcal{Q}}}(W^{2})}.
\end{eqnarray}

\subsection{4. Results}
$\mathbf{4.1:~ Charm~ and~ Beauty}$\\

In Refs.[1] and [2], the reduced cross sections and structure
functions of the charm and beauty quarks in center-of-mass
energies $\sqrt{s}=319~\mathrm{GeV}$ and
$\sqrt{s}=318~\mathrm{GeV}$ can be observed respectively. The mass
of the charm and beauty quarks  set to $m_{c}=1.5~\mathrm{GeV}$
and $m_{b}=4.75~\mathrm{GeV}$ respectively. The extracted values
of $\frac{F_{L}^{c\overline{c}}}{F_{2}^{c\overline{c}}}$ and
$\frac{F_{L}^{b\overline{b}}}{F_{2}^{b\overline{b}}}$ into HERA
data in Refs.[1] and [2] are given in figures 1 and 2
respectively. In these figures the ratio of structure functions
for the charm and beauty quarks is plotted into a wide range of
the invariant of mass $W^{2}$. Calculations allowing the invariant
mass $W^{2}$ to vary in the interval ($5000<W^{2}<60000~
\mathrm{GeV}^{2}$) when $x$ and $Q^{2}$ varies according to HERA
data. We observed that these ratio of structure functions are
statistically very scattered. However, according to the Eqs.(19),
the linear fit of these HERA data indicates  the difference
between intercepts for charm and beauty quarks production. The
results of this linear fit are given in Table 2. Our belief is
that the behavior of
$\lambda_{2}^{\mathcal{Q}\overline{\mathcal{Q}}}$ corresponds to
hard pomeron behavior. Based on this,
$\lambda_{L}^{\mathcal{Q}\overline{\mathcal{Q}}}$ intercept can be
determined based on HERA data. Nevertheless the data have the
striking property that the ratio of structure functions behave by
the following form
\begin{eqnarray}
\frac{F_{L}^{\mathcal{Q}\overline{\mathcal{Q}}}(W^{2})}{F_{2}^{\mathcal{Q}\overline{\mathcal{Q}}}(W^{2})}.
\frac{f_{2}^{\mathcal{Q}\overline{\mathcal{Q}}}}{f_{L}^{\mathcal{Q}\overline{\mathcal{Q}}}}=
(W^{2})^{\Delta{\lambda_{L2}^{\mathcal{Q}\overline{\mathcal{Q}}}}},
\end{eqnarray}
where $F_{2,L}^{\mathcal{Q}\overline{\mathcal{Q}}}(W^{2}){
\simeq}f_{2,L}^{\mathcal{Q}\overline{\mathcal{Q}}}.
(W^{2})^{\lambda_{2,L}}$. Figure 3 shows this ratio (i.e.,
Eq.(21)) according to Table II and shows that the importance of
measuring the longitudinal structure function for beauty quark  is
not less than charm quark. In this figure,
$\Delta\lambda=\lambda_{L}-\lambda_{2}$ is obtained from a linear
fit to the heavy quark structure functions into the invariant
center-of- mass energy. Because the heavy-quark longitudinal
structure function data scatter is high for H1 and ZEUS (according
to Figs.1 and 2), therefore the linear fit of the data shows a
noticeable difference in this figure. However, these fits can give
comparable results below.
 According to Eq.(14), our results for
the ratio
$F_{L}^{\mathcal{Q}\overline{\mathcal{Q}}}/F_{2}^{\mathcal{Q}\overline{\mathcal{Q}}}$
are described in Fig.4 for charm and beauty quarks in a wide range
of the invariant mass $W^{2}$. The mass of charm and beauty quarks
are defined according to Table III. In this figure, our results at
$Q^{2}=60$ and $80~\mathrm{GeV}^{2}$ are compared with HERA data
in Refs.[1,2]. These results are comparable with HERA data. As we
see in this figure (i.e., Fig.4), the value of this ratio is
almost constant in a wide rang of the invariant mass. This ratio
for the charm quark at both $Q^{2}$ values is almost $0.21$ and
for the beauty quark is almost $0.13-0.15$. This conclusion for
charm quark is close to the results Refs.[17,23,29]. In Fig.5 we
present the ratio
$F_{L}^{\mathcal{Q}\overline{\mathcal{Q}}}/F_{2}^{\mathcal{Q}\overline{\mathcal{Q}}}$
for charm and beauty quarks as a function of $Q^{2}$ at
$W=200~\mathrm{GeV}$. We can see that obtained results are in
agreement with others. The maximum value for charm and beauty
ratios is the same and is equal to $\simeq 0.21$ in a wide range
of $Q^{2}$ values. This maximum value shifts to larger $Q^{2}$
values for beauty quark. Indeed it shifts from
$Q^{2}{\simeq}60~\mathrm{GeV}^{2}$ for charm to
$Q^{2}{\simeq}800~\mathrm{GeV}^{2}$ for beauty quark. In these
calculations, the errors are due to calculation errors related to
the charm and beauty quarks mass and the gluon intercept. Indeed,
the average between the hard-pomeron and color dipole models for
the gluon exponent is assumed that  $\lambda_{g}=0.4{\pm}0.1$. In
this figure we also compared our results for the ratio of
structure functions for charm and beauty with HERA data [1,2].
Although errors related to the experimental data are not
available, but the comparison of these results with HERA data are
very good. It should be noted that the data collected from HERA is
related to $175<W<225~\mathrm{GeV}$,
and this is the reason for the error between our results (at $W=200~\mathrm{GeV}$) and HERA data.\\
Now we focus attention on the energy shift from HERA to LHeC. LHeC
data will also allow us to increase our knowledge of heavy flavour
structure functions [30]. Due to the increase in  center-of-mass
energy in new colliders, the LHeC provides data on charm and
beauty structure functions extending over nearly 5 and 6 orders of
magnitude in $x$ and $Q^{2}$ respectively [9]. According to the
predicted energy range for LHeC, the center-of-mass energy changes
as follows, $10^{2}{\leq}W^{2}{\leq}10^{6}~\mathrm{GeV}^{2}$. In
Fig.6, phenomenological predictions of the charm and beauty
structure functions are determined in center-of-mass energy
$\sqrt{s}=1.3~\mathrm{TeV}$. We can see that as the energy
increases, the maximum values for these ratios are still the same
as $\simeq 0.21$. But the amount of $Q^{2}$ values increases
slightly, as the maximum  ratio value
$F_{L}^{c\overline{c}}/F_{2}^{c\overline{c}}$ is in the order
$\mathcal{O}(100~\mathrm{GeV}^{2})$ and the maximum ratio value
$F_{L}^{b\overline{b}}/F_{2}^{b\overline{b}}$ is in the order
$\mathcal{O}(1000~\mathrm{GeV}^{2})$. In this figure (i.e.,
Fig.6), we show the $Q^{2}$ dependents of the heavy-quarks
structure functions evaluated at NLO analysis.\\

$\mathbf{4.2:~ Top}$\\

Top-quark pairs can be produced the LHeC and FCC-eh from
$\gamma^{*}p{\rightarrow}t\overline{t}$ reactions. Deep inelastic
scattering measurements at LHeC and FCC-eh will allow the
determination of the top distribution function. The production of
top quarks in ep collisions  at LHeC and FCC-eh can be  provided a
stringent test of new physics at ultra-high energy (UHE), which
the center-of-mass energies are $1.3~\mathrm{TeV}$ and
$3.5~\mathrm{TeV}$ respectively. One of the QCD corrections in top
quarks production is the QCD coupling $\alpha_{s}(\mu^{2})$ (
$\mu^{2}{\propto}\beta^{2}\widehat{s}$ where $\beta$ is the
$\mathcal{Q}\overline{\mathcal{Q}}$ relative velocity and
$\widehat{s}=4m_{t}^{2}$) [30]. The running coupling constant at
LHeC is the H1 result at NNLO analysis with $0.2\%$ uncertainty
from the LHeC and $0.1\%$ uncertainty when combined with HERA data
[8,9,10,31]. Here we used the active flavor number $n_{f}=6$ in
the running of $\alpha_{s}$ [9,10]. Because the top threshold is
high enough, therefore the range of inelasticity changes is very
important in determining the ratio
$F_{L}^{t\overline{t}}/F_{2}^{t\overline{t}}$. For
$Q^{2}=40000-50000~\mathrm{GeV}^{2}$ where $Q^{2}>m_{t}^{2}$
($m_{t}=172{\pm}0.5~\mathrm{GeV}$), the area available in the
further colliders  covers the inelasticity from $0.300<y<0.800$
and $0.080<y<0.850$ at LHeC and FCC-eh respectively. The Wilson
coefficient functions are experiencing a slow rescaling by
replacing $z{\rightarrow}\chi$. This changes the integration range
of $z$ in the convolutions to
$x(1+\frac{4m_{t}^{2}}{Q^{2}}){\leq}z{\leq}1$ with the Bjorken
variable $x$ [15]. In the following we will discuss the ratio
$F_{L}^{t\overline{t}}/F_{2}^{t\overline{t}}$ in the
$t\overline{t}$ production at the LHeC and  FCC-eh. In Fig.7 we
present this ratio for $Q^{2}=40000$ and $50000~\mathrm{GeV}^{2}$
 as a function of $W^{2}$. Because $0<y{\leq}1$ then the range of $x$ changes is very
limited. Notice that the large inelasticity is only for scattered
electron energies much smaller than the electron beam energy
(i.e., $E'_{e}{\ll}E_{e}$ and $y=1-E'_{e}/E_{e}$). In this region
which $E'_{e}$ is small, the electromagnetic and hadronic
backgrounds are important [9]. The maximum value for the ratio
$F_{L}^{t\overline{t}}/F_{2}^{t\overline{t}}$ with respect to the
inelasticities (i.e., $0.300<y<0.800$ and $0.080<y<0.850$ at LHeC
and FCC-eh, respectively)  is almost $\simeq 0.1$. Of course, this
value for the ratio $F_{L}^{t\overline{t}}/F_{2}^{t\overline{t}}$
also increases to $\simeq 0.21$, when both $Q^{2}$ and $x$ values
increase. In this regard, the $Q^{2}$ value must be of order
$\mathcal{O}(100000~\mathrm{GeV}^{2})$ at LHeC and of order
$\mathcal{O}(1000000~\mathrm{GeV}^{2})$ at FCC-eh and the $x$
value of order $\mathcal{O}(0.1)$. All these predictions can be
seen in Fig.8. In this figure, the ratio of structure functions
 results for charm, beauty  and top pair production are visible in
the range of energies available for future accelerators. However,
this range of $x$ and $Q^{2}$ will be seen in future
accelerations.\\
For the calculations presented, we considered the mass error and
the gluon exponent error. In all figures, bandwidth errors are
included. In Fig.9, we considered the effect of the
renormalization/factorization scale uncertainty in the ratio
$F_{L}/F_{2}$ for charm and beauty due to the LHeC center of mass
energy. The left and right panels of Fig.9 show the ratio
$F_{L}/F_{2}$ for charm and beauty as a function of $Q^{2}$ for
$x=0.001$ respectively. Our NLO results for $\mu_{c(b)}=2m_{c(b)}$
and $\mu_{c(b)}=\sqrt{4m^{2}_{c(b)}+Q^{2}}$ are presented in this
figure, where they are compared with the quantities represented in
Refs.[17](A.~Y.~Illarionov,B.~A.~Kniehl and A.~V.~Kotikov,
Phys.Lett. B ${\bf 663}$, 66 (2008)) and [23](N.Ya.Ivanov, and
B.A.Kniehl, Eur.Phys.J.C$\textbf{59}$, 647(2009)). In both
references, the ratios are independent of the choice of the gluon
distribution. These approaches based on perturbative QCD and
$k_{T}$ factorization gives similar predictions for the ratio of
the heavy-quark structure functions. One of them (N.Ya.Ivanov, and
B.A.Kniehl, Eur.Phys.J.C$\textbf{59}$, 647(2009)) provide
analytical result for the ratio of structure functions for
arbitrary values of the parameter $\lambda_{g}$ in terms of the
Gauss hypergeometric function. Those consider compact formulae for
the ratio in two particular cases $\lambda_{g}=0$ and
$\lambda_{g}=0.5$. The simplest case leads to a non-singular
behavior at small $x$ for the structure functions and another one
(i.e., $\lambda_{g}=0.5$) originates from the BFKL resummation of
the leading powers of $\ln(1/x)$. In another reference, authors
provide compact formulas for the ratio of structure functions with
respect to the Mellin transform. In the following for low and
moderate $Q^{2}$ one should take into account quark mass. So we
replace $x{\rightarrow}(1+\frac{4m^{2}_{\mathcal{Q}}}{Q^{2}})x$ in
the formula for the ratio of heavy-quark structure functions.
 The behavior of the ratios are
much less sensitive to the choice of scale $\mu$ at low and
moderate values of $Q^{2}$, as seen by comparing the corresponding
curves in the two figures. For $Q^{2}{\gg}4m^{2}_{c(b)}$ the NLO
predictions exhibit an appreciable scale dependence. One can see
that all the considered NLO predictions agree with the literature
results with a good
accuracy. \\

\subsection{5. Summary and Conclusion}

We presented the  ratio
$F_{L}^{\mathcal{Q}\overline{\mathcal{Q}}}/F_{2}^{\mathcal{Q}\overline{\mathcal{Q}}}$
for charm and beauty pair production with respect to HERA data.
The behavior of these ratios are in good agreement in comparison
with experimental data  in a wide range of $W^{2}$ values. Indeed
a power-law behavior for the ratio of structure functions for
heavy quark pair production is predicted. Results as a function of
$Q^{2}$ for an invariant constant value are in agreement with
those extracted in literature in the framework of perturbative
QCD. Then we have studied the production of heavy-pair quarks in
new electron proton collisions (i.e., LHeC and FCC-eh). The ratio
of charm and beauty structure functions are studied in center of
mass energy $\sqrt{s}=1.3~\mathrm{TeV}$ which proposed at LHeC
collider. For the top quark pair production, which will be one
kind of important production channel at LHeC and FCC-eh, the ratio
of structure functions determined and compared together with
respect to the center-of-mass energies in new colliders. The
results of numerical calculations for heavy quarks in LHeC and
FCC-eh are available with respect to the inelasticity defined in
accordance with the center of mass energies. These results
highlight the importance of measuring the longitudinal structure
function in the production of heavy quarks in the future with
respect to the energies available in the new accelerators.\\

\subsection{ACKNOWLEDGMENTS}
The author is thankful to the Razi University for financial
support of this project. The author is also especially grateful to
Max Klein for carefully reading the manuscript and fruitful
discussions and thanks H.Khanpour for interesting
 and useful discussions. \\

\begin{table}[h]
\centering \caption{Illustration of the kinematic covrage of
simulated  heavy quark structure functions used in LHeC studies.
The number of pseudo-data points, $N_{dat}$ and the integrated
luminosity, $ \mathcal{L}_{\mathrm{int}}[\mathrm{ab}^{-1}] $, are
shown in this table [9,14].}\label{table:table2}
\begin{minipage}{\linewidth}
\renewcommand{\thefootnote}{\thempfootnote}
\centering
\begin{tabular}{|c|c|c|c|c|c|} \hline\noalign{\smallskip}  $ \mathrm{Observable} $ & $ E_{e} $ & $ E_{p}$ & $ \mathrm{Kinematics} $ & $ \mathrm{N}_{\mathrm{dat}} $ & $ \mathcal{L}_{\mathrm{int}}[\mathrm{ab}^{-1}] $ \\
\hline\noalign{\smallskip}
$F_{2}^{c,NC}(e^{-}p)$ & 60 GeV & 7 TeV & $7{\times}10^{-6}{\leq}x{\leq}0.3, 4{\leq}Q^{2}{\leq}2{\times}10^{5} \mathrm{GeV}^{2}$& 111& 1.0 \\
\hline
$F_{2}^{b,NC}(e^{-}p)$ & 60 GeV & 7 TeV & $3{\times}10^{-5}{\leq}x{\leq}0.3, 32{\leq}Q^{2}{\leq}2{\times}10^{5} \mathrm{GeV}^{2} $& 77& 1.0 \\
\hline\noalign{\smallskip}
\end{tabular}
\end{minipage}
\end{table}
\begin{table}[h]
\centering
\caption{$\Delta{\lambda}_{L2}^{\mathcal{Q}\overline{\mathcal{Q}}}=\lambda_{L}^{\mathcal{Q}\overline{\mathcal{Q}}}-
\lambda_{2}^{\mathcal{Q}\overline{\mathcal{Q}}}$ determined for
heavy-quark pair production (HQPP) according to HERA collected
data in Refs.[1,2].}\label{table:table2}
\begin{minipage}{\linewidth}
\renewcommand{\thefootnote}{\thempfootnote}
\centering
\begin{tabular}{|c|c|c|} \hline\noalign{\smallskip}  $ HQPP $ & $ \Delta{\lambda}_{L2}^{\mathcal{Q}\overline{\mathcal{Q}}} $ & Coll.Data  \\
\hline\noalign{\smallskip}
$c\overline{c}$ & ${\sim} -0.68$ & Ref.[1]  \\
\hline
$c\overline{c}$ & ${\sim} -0.77$ & Ref.[2]  \\
\hline
$b\overline{b}$ & ${\sim} -0.24$ & Ref.[1]  \\
\hline
$b\overline{b}$ & ${\sim} -0.41$ & Ref.[2]  \\
\hline\noalign{\smallskip}
\end{tabular}
\end{minipage}
\end{table}
\begin{table}[h]
\centering \caption{Heavy quarks mass with respect to the
statistical and systematic uncertainties [3].}\label{table:table2}
\begin{minipage}{\linewidth}
\renewcommand{\thefootnote}{\thempfootnote}
\centering
\begin{tabular}{|c|c|c|c|c|} \hline\noalign{\smallskip}  Quark  & Mass & exp/fit & Model & Parameterization \\
\hline\noalign{\smallskip}
c & 1.290 GeV & $^{\mathbf{+0.046}}_{\mathbf{-0.041}}$ &  $^{\mathbf{+0.062}}_{\mathbf{-0.014}}$ &  $^{\mathbf{+0.003}}_{\mathbf{-0.031}}$\\
\hline
b & 4.049 GeV &  $^{\mathbf{+0.104}}_{\mathbf{-0.109}}$ & $^{\mathbf{+0.090}}_{\mathbf{-0.032}}$ & $^{\mathbf{+0.001}}_{\mathbf{-0.031}}$ \\
\hline\noalign{\smallskip}
\end{tabular}
\end{minipage}
\end{table}

\newpage{
\section{References}
1. F.D.Aaron et al., [H1 Collaboration], Eur.Phys.J.C{\bf65}, 89
(2010).\\
2. H.Abramowicz et al., [ZEUS Collaboration], JHEP{\bf09}, 127
(2014).\\
3. H.Abramowicz et al., [H1 and ZEUS Collaboration],
Eur.Phys.J.C{\bf78}, 473(2018).\\
4. H.Abramowicz et al., [H1 and ZEUS Collaboration], DESY-12-172,
arXiv:1211.1182v2 [hep-ex](2012); H.Abramowicz et al., [ ZEUS
Collaboration], Eur.Phys.J.C{\bf69}, 347 (2010).\\
5.  E. Laenen et al., Phys. Lett. B{\bf291}, 325 (1992); E. Laenen
et al., Nucl. Phys. B{\bf392}, 162 (1993); S. Riemersma, J.
Smith,W.L. van Neerven, Phys. Lett. B{\bf347}, 143 (1995); S.
Alekhin et al., Phys. Rev.D{\bf81}, 014032 (2010); S. Alekhin, S.
Moch. arXiv:1107.0469 (2001); S. Alekhin, J.
Bl$\ddot{\mathrm{u}}$mlein, S. Moch, Phys. Rev. D{\bf86}, 054009
(2012); S. Alekhin et al., Phys. Rev. D{\bf96}, 014011 (2017); S.
Alekhin et al., arXiv:0908.3128 [hep-ph](2009); M.
Gl$\ddot{\mathrm{u}}$ck et al., Phys. Lett. B{\bf664}, 133 (2008);
H.L. Lai et al., Phys. Rev. D{\bf82}, 074024 (2010); A.D. Martin
et al., Eur. Phys. J. C{\bf70}, 51 (2010); S. Alekhin and S. Moch,
Phys. Lett. B{\bf699}, 345 (2011).\\
6. S. Forte et al., Nucl. Phys. B{\bf834}, 116 (2010); R.D. Ball
et al. [NNPDF Collaboration], Nucl. Phys. B{\bf849}, 296 (2011);
R.D. Ball et al. [NNPDF Collaboration], Nucl. Phys. B{\bf855}, 153
(2012); R.D.
Ball et al. arXiv:1710.05935 (2017).\\
7. R.Thorne, Phys.Rev.D{\bf73}, 054019 (2006); R.Thorne,
Phys.Rev.D{\bf86}, 074017 (2012).\\
8. M.Klein,  arXiv :1802.04317 [hep-ph] (2018); M.Klein, Ann.Phys.{\bf528}, 138(2016); N.Armesto et al., Phys.Rev.D{\bf100}, 074022(2019).\\
9.  J.Abelleira Fernandez et al., [LHeC Collaboration], J.Phys.G{\bf39}, 075001(2012);
 P.Agostini et al., [LHeC Collaboration and FCC-he Study Group], CERN-ACC-Note
 -2020-002, arXiv:2007.14491 [hep-ex](2020).\\
10. A. Abada et al., [FCC Collaboration], Eur.Phys.J.C{\bf 79}, 474(2019).\\
11. B.W.Harris and J.Smith, Phys.Rev.D{\bf57}, 2806 (1998).\\
12. LHeC data, http://hep.ph.liv.ac.uk/$^{\sim}$mklein/lhecdata/.\\
13. Heavy Q data, http://hep.ph.liv.ac.uk/$^{\sim}$mklein/heavydata/.\\
14. R. Abdul khalek, et al., SciPost Phys.{\bf7}, 051 (2019).\\
15. S. Alekhin, J.Bl$\ddot{\mathrm{u}}$mlein and  S. Moch,
arXiv:2006.07032 [hepph] (2020).\\
16. Wu-Ki Tung, S.Kretzer and C.Schmidt, J.Phys.G{\bf28},
983(2002).\\
 17. U.Baur and J.J.Van Der Bij, Nucl.Phys.B{\bf304},
451(1988); A.~Y.~Illarionov,B.~A.~Kniehl and A.~V.~Kotikov,
Phys.Lett. B {\bf 663}, 66 (2008); S. Catani and F. Hautmann,
Nucl. Phys. B \textbf{427}, 475(1994); S. Riemersma, J. Smith and
W. L. van Neerven, Phys. Lett. B \textbf{347}, 143(1995);
 F.~P.~Wi$\ss$brock, DESY-THESIS-2015-040 (2015);  B.W. Harris and J. Smith, Nucl.Phys. B{\bf452},
 109(1995).\\
18. I.P.Ivanov and N.N.Nikolaev, arXiv:hep-ph/0004206 (2000);
 J.Lan et al., Phys.Rev. D{\bf102}, 014020 (2020);
 J.Lan et al.,  Phys.Rev. D{\bf101}, 034024 (2020); C.Mondal, Eur.Phys.J.C{\bf76}, 74(2016).\\
19. A.Donnachie and P.V.Landshoff, Z.Phys.C \textbf{61},
139(1994); Phys.Lett.B \textbf{518}, 63(2001); Phys.Lett.B
\textbf{533}, 277(2002); Phys.Lett.B \textbf{470}, 243(1999);
Phys.Lett.B \textbf{550}, 160(2002); R.D.Ball and P.V.landshoff,
J.Phys.G\textbf{26}, 672(2000); P.V.landshoff,
arXiv:hep-ph/0203084 (2002).\\
 20. N.N.Nikolaev
and V.R.Zoller, Phys.Lett. B\textbf{509}, 283(2001); A. V.
Kotikov, A. V. Lipatov, G. Parente and N. P. Zotov Eur.
Phys.J.C{\bf 26}, 51(2002).\\
21. C.Adloff et al. [H1 Collaboration], Phys.Lett.B{\bf520},
183(2001); R.D.Ball et al., Eur.Phys.J.C{\bf76}, 383(2016);
 G.R.Boroun, Eur.Phys.J.Plus {\bf135}, 68(2020); B.Rezaei and G.R.Boroun, Eur.Phys.J.A{\bf55}, 66(2019).\\
 22. J.Bartels et al., Acta.Phys.Polon.B{\bf33}, 2853(2002);
J.Bartels et al., Phys.Rev.D{\bf66}, 014001(2002);
 E.Iancu et al., Phys.Lett.B{\bf590}, 199(2004); A.M.Stasto et al., Phys.Rev.Lett.{\bf86}, 596 (2001).\\
23. J.Lan et al., arXiv [nucl-th]:1911.11676 (2019); N.N.Nikolaev
and V.R.Zoller, Phys.Atom.Nucl\textbf{73}, 672(2010);
N.N.Nikolaev, J.Speth and V.R.Zoller, Phys.Lett.B\textbf{473},
157(2000); R.Fiore, N.N.Nikolaev and V.R.Zoller, JETP
Lett\textbf{90}, 319(2009); A. Y. Illarionov and A. V. Kotikov,
Phys.Atom.Nucl. {\bf75}, 1234 (2012); N.Ya.Ivanov, and B.A.Kniehl,
Eur.Phys.J.C\textbf{59}, 647(2009);  N.Ya.Ivanov,
Nucl.Phys.B\textbf{814}, 142(2009); J.Bl$\ddot{\mathrm{u}}$mlein,
et.al.,
Nucl.Phys.B\textbf{755}, 272(2006); A.Kotikov, arXiv:1212.3733 [hep-ph] (2012).\\
 24. G.R.Boroun and B.Rezaei,
Int.J.Mod.Phys.E{\bf24}, 1550063(2015); G.R.Boroun and B.Rezaei,
Nucl.Phys.A{\bf929}, 119(2014); G.R.Boroun, Nucl.Phys.B{\bf884},
684(2014); G.R.Boroun and B.Rezaei, EPL{\bf100}, 41001(2012);
G.R.Boroun and B.Rezaei, J.Exp.Theor.Phys.{\bf115}, 427(2012);
G.R.Boroun and B.Rezaei, Nucl.Phys.B{\bf857}, 143(2012).\\
25. H.Sun, POS(DIS2018)186; H.Sun, arXiv
[hep-ph]:1710.06260(2017); Z.Zhang, arXiv[hep-ex]: 1511.05399
(2015).\\
26. Ch. Schwanenberger, LHeC worksope, CERN/Chavannes-de-Bogis
(2015);
 Ch. Schwanenberger, POS(DIS2017)464; A.O.Bouzas and F.Larios, Journal of Physics:
 Conference Series {\bf 651}, 012004(2015).\\
27. G.R.Boroun, Phys.Lett.B{\bf744}, 142(2015); G.R.Boroun,
Phys.Lett.B{\bf741}, 197(2015); G.R.Boroun, Chin.Phys.C{\bf414},
013104(2017); G.R.Boroun, PEPAN Lett.{\bf15}, 387(2018);
G.R.Boroun
, B.Rezaei and S.Heidari, Int.J.Mod.Phys.A{\bf32}, 1750197(2017);
 B.Rezaei and G.R.Boroun, EPL{\bf130}, 51002(2020); H.Khanpour, Nucl.Phys.B{\bf
 958}, 115141(2020).\\
28. A.O.Bouzas and F.Larios, Phys.Rev.D{\bf88}, 094007(2013).\\
29. V.P.Goncalves and M.V.T.Machado, Phys.Rev.Lett.{\bf91},
202002(2003); B.Rezaei and G.R.Boroun, Phys.Rev.C{\bf101},
045202(2020); G.R.Boroun and B.Rezaei, Nucl.Phys.A{\bf990},
244(2019).\\
30. Amanda Cooper-Sarkar, arXiv: 1310.0662 [hep-ph](2013); Stanley
J.Brodsky, arXiv: 1106.5820 [hep-ph](2011).\\
31. F.Demartin et al., Phys.Rev.D{\bf82}, 014002(2010).\\

}

\newpage{
\begin{figure}
\includegraphics[width=0.55\textwidth]{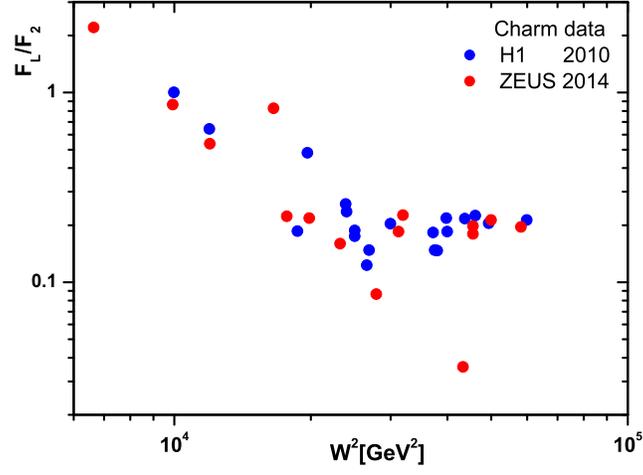}
\caption{The ratio of charm structure functions, with respect to
HERA data (H1 2010 [1] and ZEUS 2014 [2]), shown as a function of
$W^{2}$ values. }\label{Fig1}
\end{figure}
\begin{figure}
\includegraphics[width=0.55\textwidth]{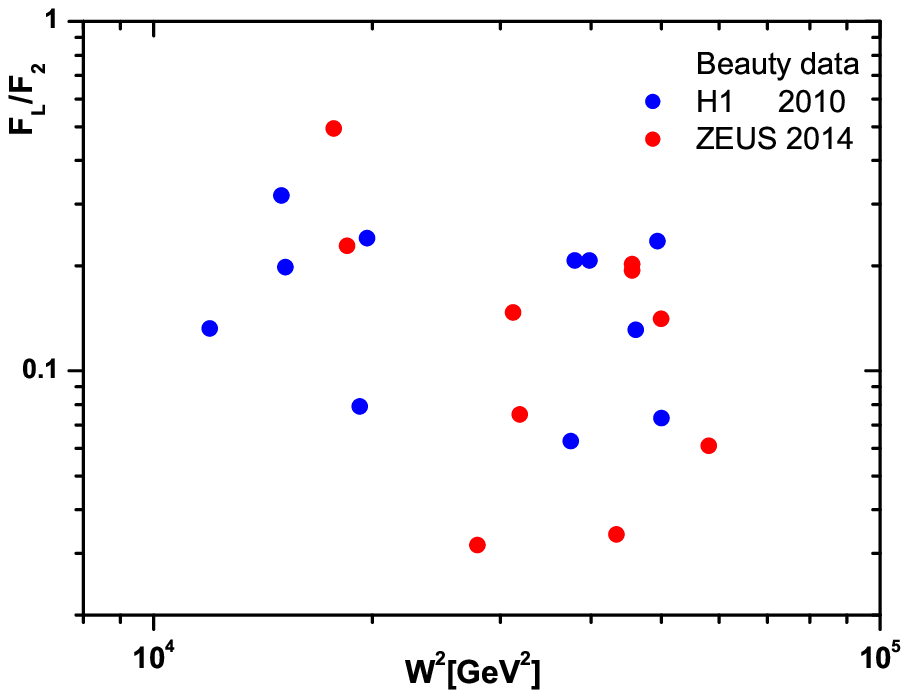}
\caption{The ratio of beauty structure functions, with respect to
HERA data (H1 2010 [1] and ZEUS 2014 [2]), shown as a function of
$W^{2}$ values. }\label{Fig1}
\end{figure}
\begin{figure}
\includegraphics[width=0.55\textwidth]{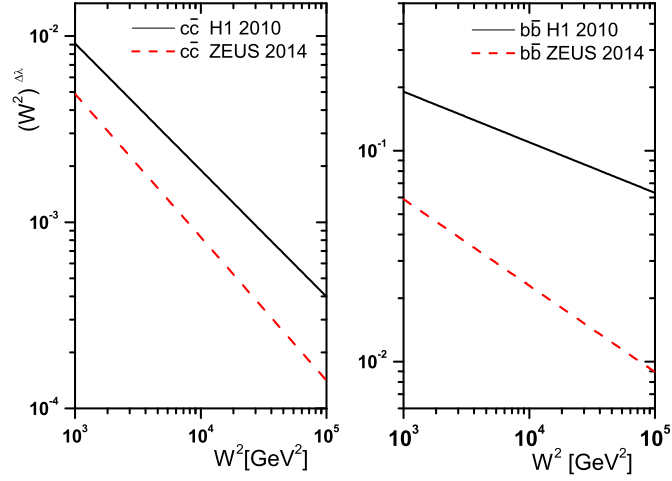}
\caption{The behavior of function $(W^{2})^{\Delta\lambda}$, for
charm and beauty with respect to HERA data (H1 2010 [1] and ZEUS
2014 [2]), shown as a function of $W^{2}$ values.
$\Delta\lambda=\lambda_{L}-\lambda_{2}$ obtained from a linear fit
to the structure functions into the invariant center-of- mass
energy.} \label{Fig1}
\end{figure}
\begin{figure}
\includegraphics[width=1\textwidth]{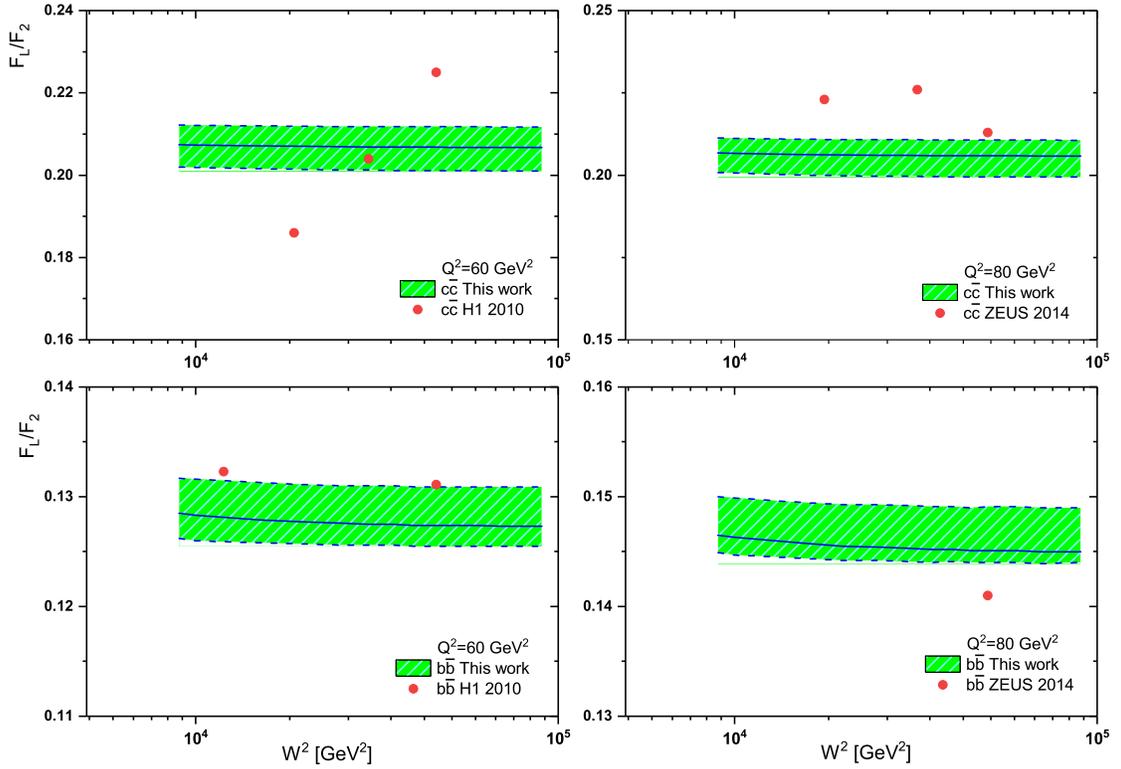}
\caption{The obtained
$F_{L}^{\mathcal{Q}\overline{\mathcal{Q}}}/F_{2}^{\mathcal{Q}\overline{\mathcal{Q}}}$
for charm and beauty pair production shown as a function of
$W^{2}$ for $Q^{2}=60$ and $80~\mathrm{GeV}^{2}$. The error bands
show the mass error and the gluon exponent error added in
prediction. The combined of HERA data [1,2] are also shown.
}\label{Fig1}
\end{figure}
\begin{figure}
\includegraphics[width=1\textwidth]{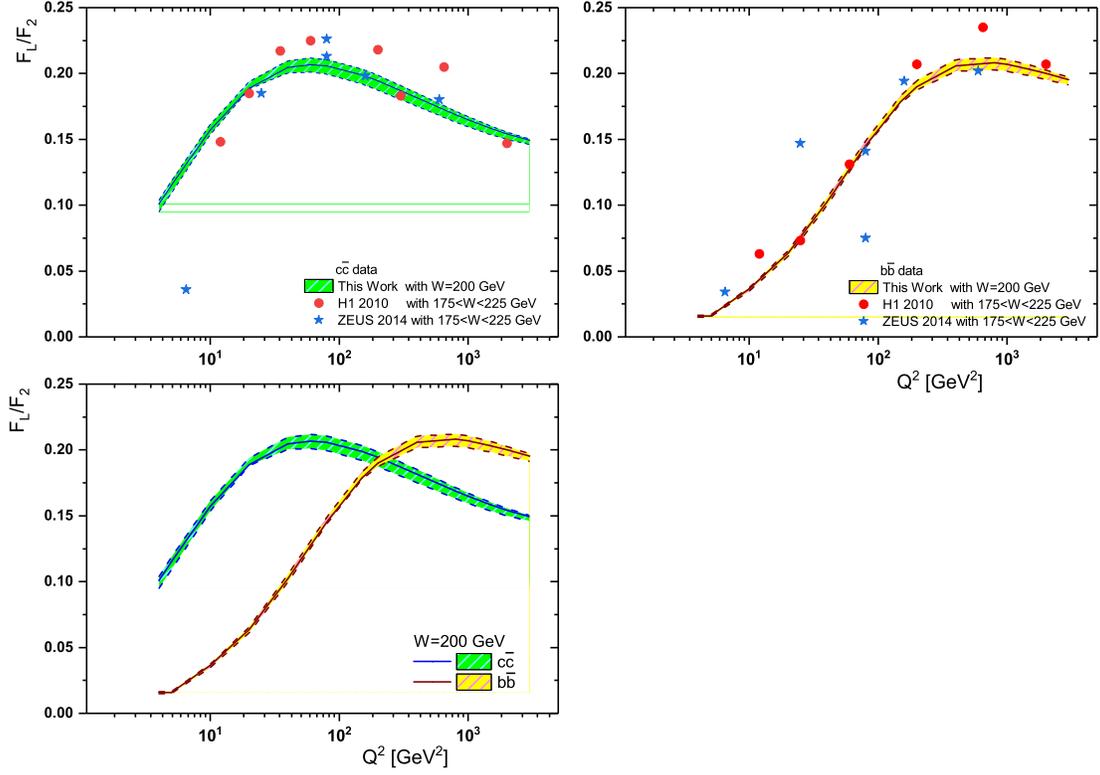}
\caption{The ratio
$F_{L}^{\mathcal{Q}\overline{\mathcal{Q}}}/F_{2}^{\mathcal{Q}\overline{\mathcal{Q}}}$
for charm and beauty pair production shown as a function of
$Q^{2}$
 for fixed $W^{2}$ value. The error bands show the mass
error and the gluon exponent error added in prediction. Our
results for charm and beauty are shown in $W=200~\mathrm{GeV}$ and
compared with H1 2010 [1] and ZEUS 2014 [2] data  in
$175<W<225~\mathrm{GeV}$. }\label{Fig1}
\end{figure}
\begin{figure}
\includegraphics[width=0.55\textwidth]{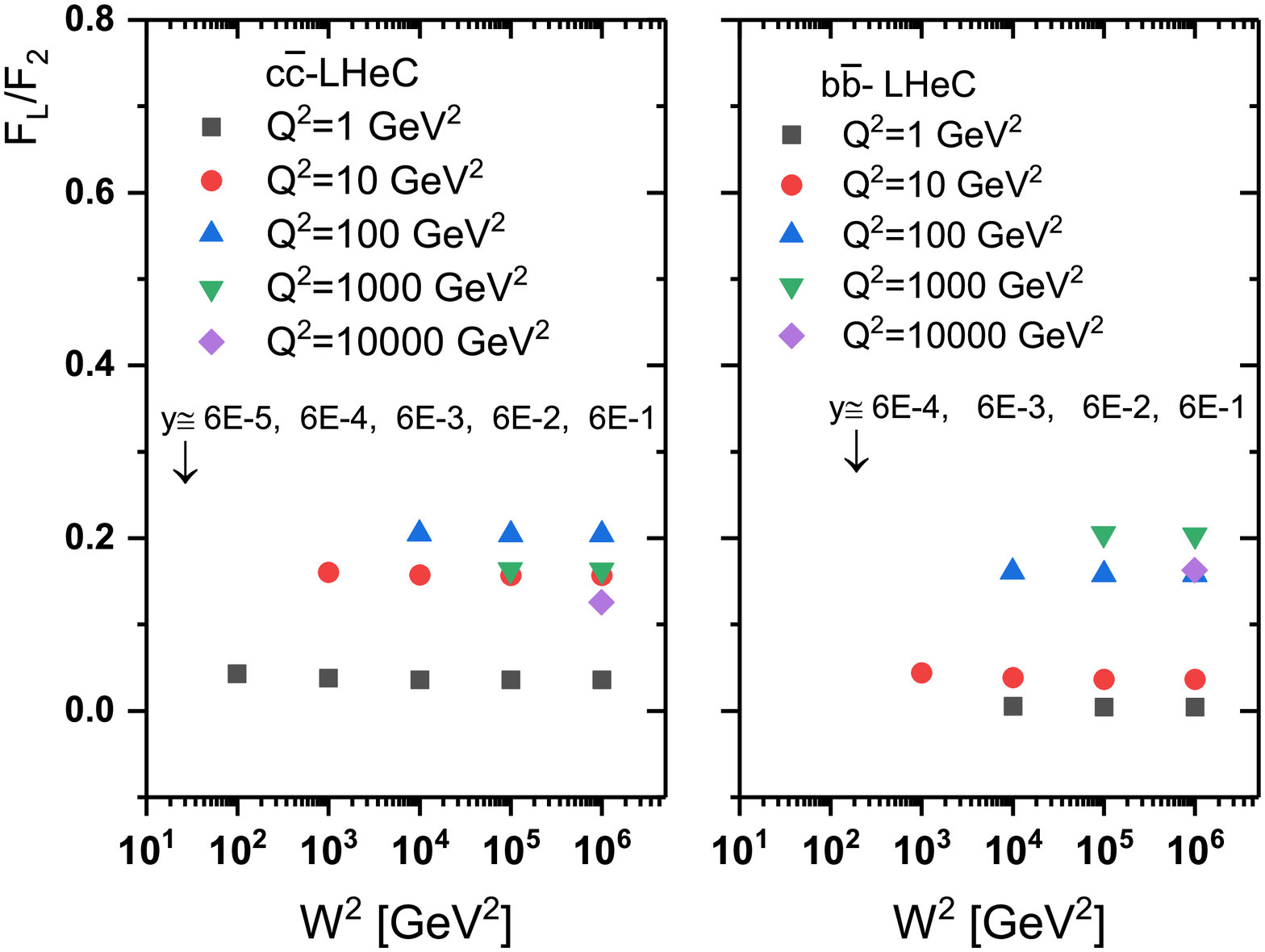}
\caption{Theoretical predictions for the ratio of charm and beauty
structure functions  at $\sqrt{s}{=}1.3~\mathrm{TeV}$ (LHeC
center-of-mass energy) shown as a function of $W^{2}$ for
different $Q^{2}$ values. The predictions for different
inelasticity $y$ values are also shown.}\label{Fig1}
\end{figure}
\begin{figure}
\includegraphics[width=0.55\textwidth]{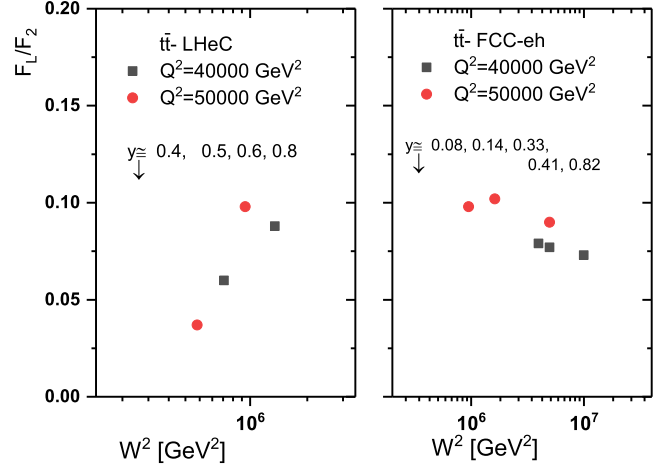}
\caption{Theoretical predictions for the ratio of top structure
functions  at $\sqrt{s}{=}1.3~\mathrm{TeV}$ (LHeC center-of-mass
energy) and at $\sqrt{s}{=}3.5~\mathrm{TeV}$ (FCC-eh
center-of-mass energy) shown as a function of $W^{2}$ for
$Q^{2}=40000$ and $50000~\mathrm{GeV}^{2}$. The predictions for
different inelasticity $y$ values are also shown.}\label{Fig1}
\end{figure}
\begin{figure}
\includegraphics[width=1\textwidth]{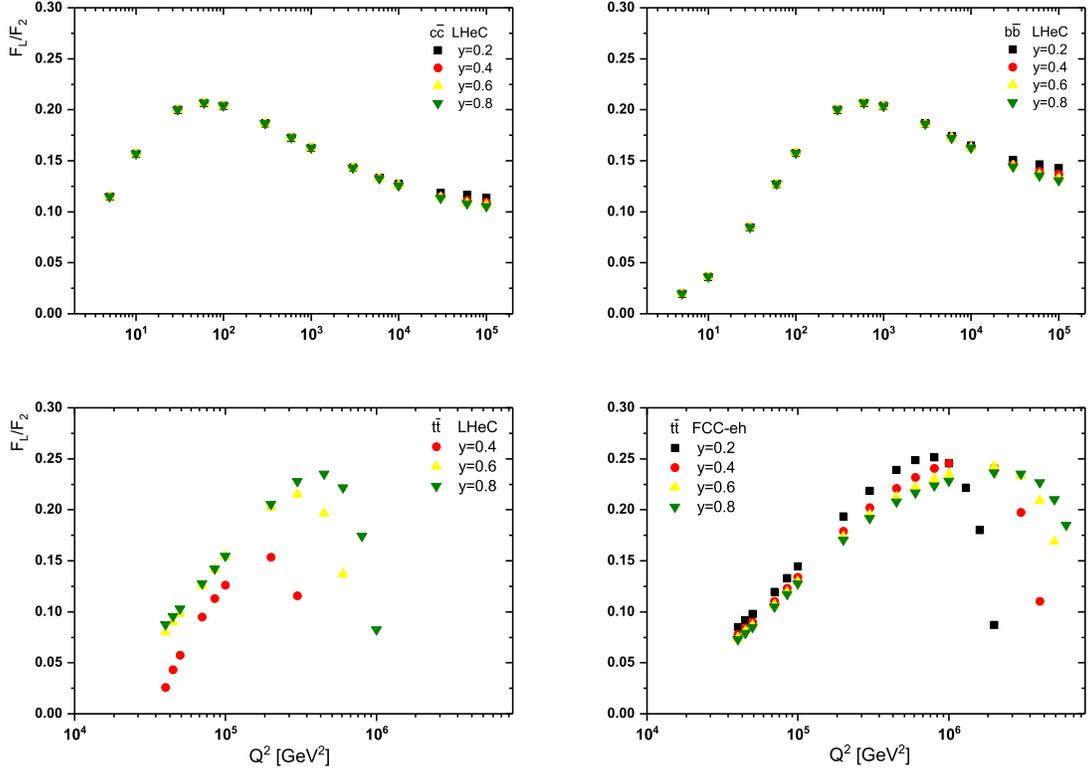}
\caption{Theoretical predictions for the ratio of charm and beauty
structure functions  at LHeC center-of-mass energy and also the
ratio of top structure functions  at LHeC and FCC-eh
center-of-mass energies  shown as a function of $Q^{2}$ for
$y=0.2$, $0.4$, $0.6$ and $0.8$.}\label{Fig1}
\end{figure}
\begin{figure}
\includegraphics[width=0.7\textwidth]{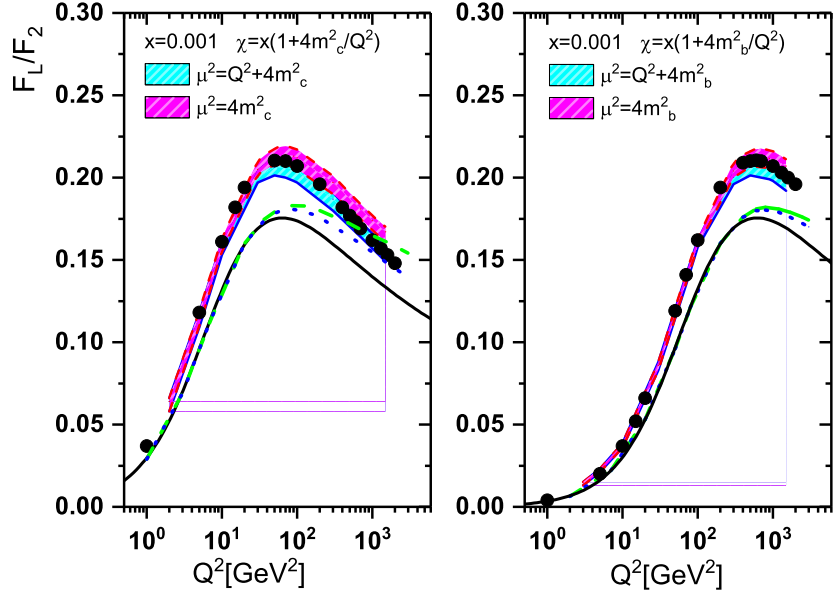}
\caption{Comparison of $F_{L}/F_{2}$ calculations for charm and
beauty vs.$Q^{2}$ with two different choices of $\mu$. These
results compared with LO (solid lines) and NLO (dash lines by
$\mu=\sqrt{4m^{2}+Q^{2}}$ and dot lines by $\mu=2m$) quantities
from Ref.[17](A.~Y.~Illarionov,B.~A.~Kniehl and A.~V.~Kotikov,
Phys.Lett. B ${\bf 663}, 66 (2008)$) and also with results
Ref.[23]( N.Ya.Ivanov, and B.A.Kniehl, Eur.Phys.J.C$\textbf{59},
647(2009)$)(black points) at $x=0.001$.}\label{Fig1}
\end{figure}
}

\end{document}